\documentclass[a4paper,11pt,twocolumn,pdftex]{nveart}
\usepackage{graphicx,amsmath,amssymb}

\begin{document}

\company{Accepted for publication in}
\journal{Astroparticle Physics}

\begin{frontmatter}
\title{On the observability of high-energy neutrinos\\
       from gamma ray bursts}

\author{Nick van Eijndhoven}
\address{Department of Physics and Astronomy, Utrecht University\\
         Princetonplein 5, NL-3584 CC Utrecht, The Netherlands\\
         Email : nickve.nl@gmail.com}

\begin{abstract}
A method is presented for the identification of high-energy neutrinos from
gamma ray bursts by means of a large-scale neutrino telescope.
The procedure makes use of a time profile stacking technique of observed neutrino
induced signals in correlation with satellite observations.
By selecting a rather wide time window, a possible difference between the arrival times
of the gamma and neutrino signals may also be identified. This might provide
insight in the particle production processes at the source.
By means of a toy model it will be demonstrated that a statistically significant
signal can be obtained with a km$^{3}$ scale neutrino telescope on a sample of 500
gamma ray bursts for a signal rate as low as 1 detectable neutrino for 3\% of the bursts.
\end{abstract}

\begin{keyword}
Neutrino astronomy,
gamma ray bursts,
neutrino telescopes.
\end{keyword}
\end{frontmatter}

\section{Introduction}
Cosmic radiation is a valuable source of information about various energetic astrophysical processes.
However, the existence of very energetic cosmic rays also raises questions such as~: how are they
accelerated and from where do they originate ?\\
A variety of possible accelerator mechanisms exists, ranging from shock waves
produced by exploding stars (supernovae) or Gamma Ray Bursts (GRBs) to supermassive
black holes with strong magnetic fields (Active Galactic Nuclei).
The current understanding is that protons and electrons are the primary particles that are
accelerated by electromagnetic fields at a cosmic accelerator site.\\
In case of a supernova event, a shock is formed by the expanding
matter envelope when it sweeps through the interstellar medium which surrounds
the exploding star. 
In such an environment stochastic processes occur which can accelerate
particles to very high energies. A detailed treatment \cite{shock} shows
that acceleration by shock waves automatically results in a power spectrum,
which is in qualitative agreement with the observations up to the 'knee' region
of the cosmic ray spectrum \cite{pdg}.\\
However, this leaves us with the question of which events can produce
the cosmic rays above the 'knee' region.
Candidates for the production of the most energetic cosmic rays are
Active Galactic Nuclei (AGN) and Gamma Ray Bursts.
The current perception is that the majority of these objects have a similar
inner engine, in which infalling matter and the likely presence of a strong
magnetic field gives rise to relativistic shock wave acceleration in two back to back jets.

\begin{figure}[htb]
\begin{center}
\includegraphics[keepaspectratio,width=5cm]{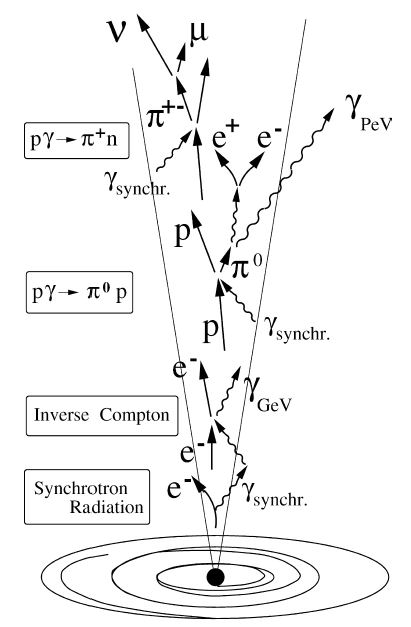}
\end{center}
\caption{Particle production in jets (courtesy C. Spiering).}
\label{fig:jet}
\end{figure}

\begin{figure}[htb]
\begin{center}
\includegraphics[keepaspectratio,width=6cm]{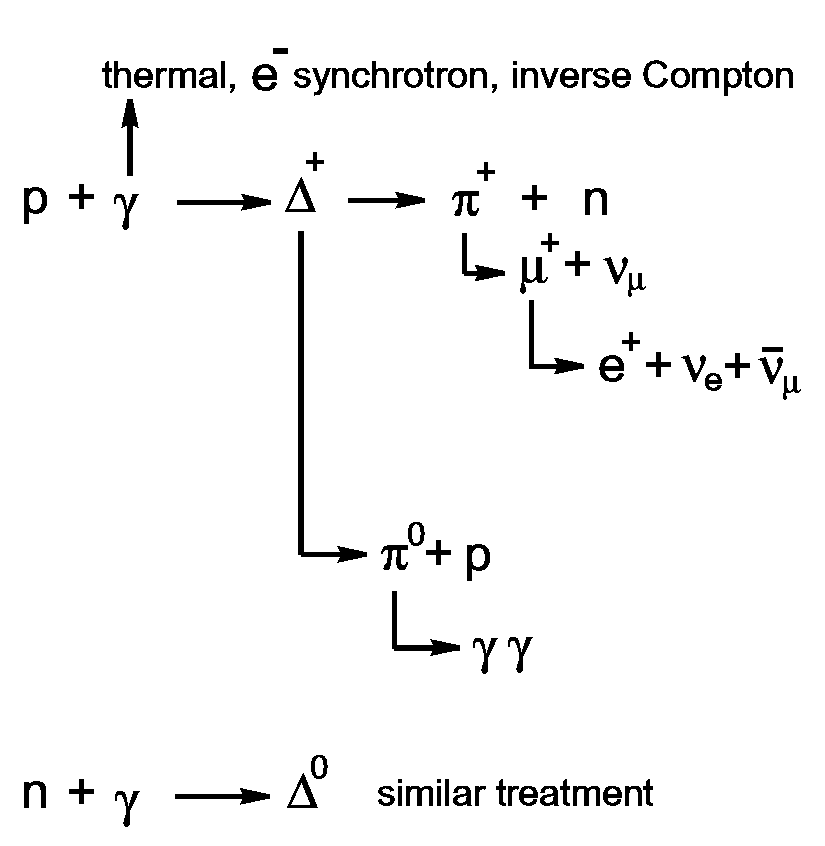}
\end{center}
\caption{Neutrino production processes.}
\label{fig:nuprod}
\end{figure}

Interactions of accelerated protons and electrons with the
ambient photons at the acceleration site give rise to very energetic
secondary particles, as shown in Fig.~\ref{fig:jet}.
In particular the $p\gamma$ interactions yield a flux of very energetic
neutrinos, as depicted in more detail in Fig.~\ref{fig:nuprod}.

In case of a proton energy of $10^{16}$~eV,
i.e. the region of the 'knee' of the cosmic ray spectrum,
the photon energy threshold for $\Delta$ production
is about 10~eV, being the ultraviolet part of the spectrum.
Since there are many of these UV photons present, the depicted hadronic processes
will take place at high rates, yielding substantial neutrino fluxes
comparable to those of ultrahigh-energy photons.\\
In the decay of the $\Delta$ resonance into a nucleon and a $\pi$ meson, the meson
obtains on average 20\% of the primary proton energy. This yields an average
neutrino energy of about 400~TeV for a primary proton energy of $10^{16}$~eV.
Detailed model calculations \cite{nuflux} predict an $E^{-2}$ powerlaw
spectrum for the produced neutrino flux.
Taking into account the fact that the atmospheric spectrum is softer \cite{pdg} and that
the neutrino cross section increases with energy \cite{pdg,nucross}, we observe that
optimal detection conditions are obtained for neutrino telescopes in an energy range
of about 10-100~TeV \cite{i3sens}.

Various attempts \cite{amagrb} have been made to identify a high-energy
neutrino flux in correlation with satellite observations of GRBs.
The performed searches for a statistical excess above the background
comprise both photon-neutrino coincidence studies and investigations
of so-called "rolling time windows".\\
However, the former will obviously fail in case there exists a significant time difference
between the arrival times of the photon and neutrino fluxes, whereas the latter
can only be succesful in case some GRBs produce multiple neutrino detections
within the corresponding time windows.
So far, no positive identifications have been reported. 

From the above it is seen that it would be preferable to use an analysis procedure that does not
require the simultaneous arrival of photons and neutrinos and which also provides
a high sensitivity in case of low signal rates.
Such a method, based on a time profile stacking technique, is presented here.

\section{The time profile stacking procedure}
In order to obtain a statistical significant result even in case of low signal
rates, a cumulative procedure as outlined below has been devised.
It is based on the generic GRB engine described in the previous section,
which implies that the arrival times of the photons and neutrinos are correlated
but are not necessarily simultaneous.

When a GRB is observed by a satellite, the trigger time $t_{grb}$ and burst location
on the sky are recorded.
Afterwards, the data of a neutrino telescope are inspected for a time interval
$[t_{grb}-\Delta t, t_{grb}+\Delta t]$ and all arrival times of upgoing muons
are recorded relative to $t_{grb}$.
Here $\Delta t$ is some predefined time margin, which is identical for all
observed bursts.
An upgoing muon is a long reconstructed track in a neutrino telescope pointing
backwards to a location in the hemisphere opposite to the detector location.
The usage of upgoing $\mu$ tracks allows reduction of the (atmospheric) background
signals in our analysis procedure, as outlined lateron.\\  
For a sample of different GRB observations, the above will result in a set of
identical time windows with upgoing $\mu$ arrival time recordings relative to the
corresponding GRB trigger time.

Stacking of all these time profiles will exhibit a uniform distribution for
background events. However, in case the data contain upgoing $\mu$
signals correlated with the GRBs, a clustering of data bins is expected.
Consequently, comparison of the stacked time profile contents
with a uniform background allows the identification of correlated signals.
Due to the cumulative character of the procedure, large statistics can be obtained
resulting in a good sensitivity even in case of low signal rates.\\
Any arrival time difference between a photon and neutrino signal poses no problem as long as
this time difference is smaller than $\Delta t$. As such, $\Delta t$ should be taken
as large as possibly allowed by the background signals. 
However, it is obvious that a spread in this photon-neutrino arrival time difference
will reduce the significance of the signal.

To address the feasibility of the procedure and to investigate the effects of the
various parameters, a toy model \footnote{See http://www.phys.uu.nl/$\sim$nick/grbmodel}
which mimics GRB induced signals as well as (atmospheric) background has been devised.
A description of this toy model and the results of the above analysis procedure
performed on the simulated data for a km$^{3}$-scale detector are presented hereafter.

\section{Signal and background generation}
Satellite observations of GRBs have shown \cite{satgrbs} that the burst locations
are homogeneously distributed over the sky.
Since our analysis procedure is based on the detection of upgoing $\mu$ tracks,
our toy model only generates GRB positions homogeneously distributed over the
hemisphere opposite to the detector location.

For each generated burst location we define the satellite trigger time to be $t_{grb} \equiv 0$
and create a time window $[-\Delta t, \Delta t]$ around it.\\
Observations with the AMANDA neutrino telescope \cite{amaupmu} show that a km$^{3}$-scale
detector will observe on average 300 upgoing muons per 24 hours due to (atmospheric) background,
homogeneously distributed over the hemisphere.
Therefore, each of the above time windows will be filled with a number of background upgoing muon
signals taken from a Poissonian distribution with an average number of
$(300/24)\cdot(2\Delta t/1 \text{ hour})$.\\
The arrival directions of these background upgoing muons are taken to be homogeneously
distributed over the hemisphere, whereas their arrival times are taken to be uniformly
distributed in the corresponding time window.
To take the detector time resolution $\sigma_{t}$ into account,
a Gaussian spread with a standard deviation $\sigma_{t}$ is introduced to the arrival times.
Finally the resulting arrival times are recorded as (background) entries in the various
corresponding time windows.
Also the angular positions of the arrival directions are recorded, after introducing a Gaussian
spread corresponding to the angular resolution $\sigma_{a}$ of the detector.
Recording of these angular positions will lateron allow reduction of the background
by correlation with the actual GRB locations.

By means of a uniform random number generator only a fraction $f$ of the generated
burst locations is selected to yield a single upgoing $\mu$ signal.
To mimic a time difference $\tau$ between the photon and neutrino burst arrival
times, the upgoing $\mu$ signal arrival time of each signal burst is taken from
a Gaussian distribution with a mean value $\tau$ and a standard deviation $\sigma_{\tau}$.
Before these signal muon arrival times are added to the corresponding time windows,
a Gaussian spread corresponding to the detector time resolution $\sigma_{t}$ is introduced.
The arrival directions of these signal upgoing muons are recorded as the locations of the
corresponding bursts, after introducing a Gaussian spread corresponding to the detector
angular resolution $\sigma_{a}$.

Introduction of realistic values for the various toy model parameters outlined above
will allow to investigate the feasibility of detecting neutrino induced signals in a large
scale neutrino telescope.
In an actual experimental data analysis effort one obviously has to account for
several additional (systematic) effects like detector stability, track reconstruction
efficiency and so on. However, these are detector specific effects and fall beyond
the scope of the present studies. 

\section{Analysis of simulated data}
The only large scale neutrino telescope currently in operation is IceCube \cite{i3perform}
and as such we use the parameters of this detector \cite{i3sens} as benchmark values
for our present studies.\\
The expected data rates for the full km$^{3}$ scale detector allow a time margin $\Delta t$
of 1 hour. This implies an average number of background signals of about 25 upgoing muons
for each individual GRB time window.
The time resolution for the reconstructed muon tracks will be of the order of the
time it takes for a muon to cross the detector volume. As such we take $\sigma_{t}=10~\mu$s
as a conservative estimate for the detector time resolution.
Experience with the analysis of the AMANDA data \cite{amaupmu,mutrack} together with detector
simulation studies \cite{i3sens} show that a realistic estimate for the angular resolution
is obtained by taking $\sigma_{a}=1^{\circ}$.\\
The remaining parameters of our toy model are related to the characteristics of the
various bursts. Based on the processes sketched in Fig.~\ref{fig:jet}, a reasonable
estimate for the possible photon-neutrino arrival time difference and its spread can
be obtained from the actual burst duration. Satellite observations \cite{satgrbs}
exhibit a mean burst duration of about 30~seconds. As such we take $\tau=30$~s
and $\sigma_{\tau}=30$~s.
As mentioned before, for evaluation of the currently presented procedure
the value of $\tau$ is actually irrelevant as long as it is smaller than the time margin $\Delta t$.\\
This leaves us with only two free parameters : the fraction $f$ of GRBs that actually induces
an upgoing muon signal and the bin size to be used for the time profiles.
In order to optimise the time bin clustering of the signals, the bin size should be taken
to be of the order of the temporal signal spread $\sigma_{\tau}$.
However, since the observed redshifts of GRBs \cite{satgrbs} exhibit a median value of $z=1.9$
with a spread of 1.3, cosmological time dilation effects have to be taken into account.
It should be noted, however, that in case both the photon and neutrino production processes
are taking place continuously throughout the jet existence, the cosmological time dilation
is already included in the observed gamma burst duration and consequently
also in the temporal signal spread $\sigma_{\tau}$. 
Nevertheless, we always account for a possible additional cosmological time dilation and take
for the time profile bin size a conservative value of $5\sigma_{\tau}$, corresponding to 150~s.\\
It should be noted here that restricting the analysis to short duration bursts
allows for smaller time bins and consequently more detailed time profile studies.\\
The fraction $f$ we keep as a free parameter in order to determine the sensitivity
of our analysis procedure for different sizes of the GRB sample.

\begin{figure}[htb]
\begin{center}
\includegraphics[keepaspectratio,width=8.5cm]{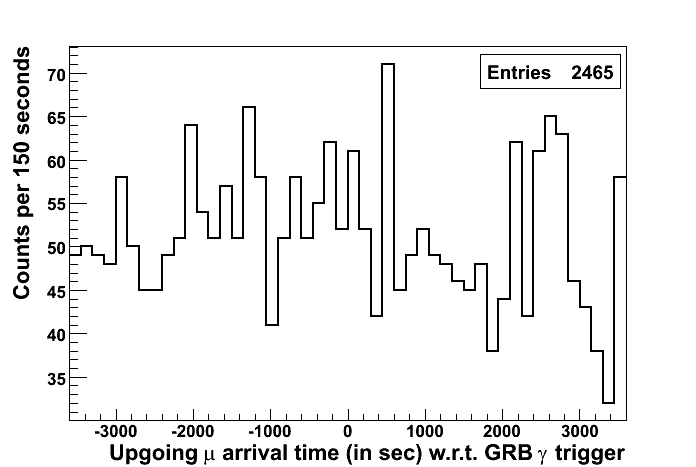}
\end{center}
\caption{Stacked time profile for 100 GRBs with $f=0.1$.
         Further details can be found in the text.}
\label{fig:tott1}
\end{figure}

For a first investigation of the performance of the procedure we generated 100 GRBs
in one hemisphere. This corresponds to about 2 years of operation of the Swift satellite \cite{swift},
which currently is the main source of GRB triggers.
All parameters were set to the values mentioned above and for the fraction $f$
we used a value of 10\% \cite{grbfrac}.
The resulting stacked time profile is shown in Fig.~\ref{fig:tott1}.

\begin{figure}[htb]
\begin{center}
\includegraphics[keepaspectratio,width=8.5cm]{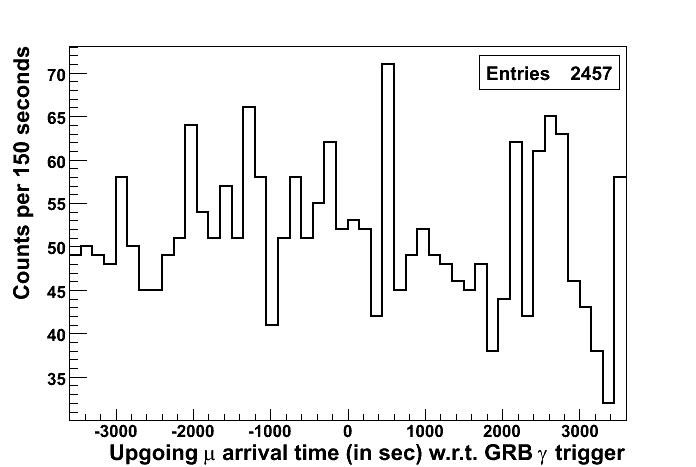}
\end{center}
\caption{Stacked time profile corresponding to the background data of Fig.~\ref{fig:tott1}.}
\label{fig:bkgt1}
\end{figure}

Since in our toy model we have access to all information, we are also able
to construct the corresponding stacked time profile from the background signals only.
This background stacked time profile is shown in Fig.~\ref{fig:bkgt1}.

Comparison of the number of entries from Fig.~\ref{fig:tott1} and Fig.~\ref{fig:bkgt1}
shows that 8 of our generated GRBs induced a signal in the stacked time window.
However, due to the presence of a large background we are not able to identify
the GRB signals on the basis of our observations of Fig.~\ref{fig:tott1} alone.

\begin{figure}[htb]
\begin{center}
\includegraphics[keepaspectratio,width=9cm]{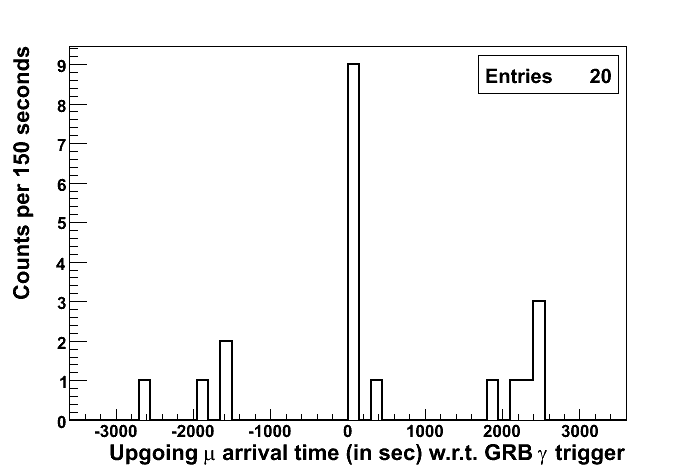}
\end{center}
\caption{Stacked time profile for 100 GRBs with $f=0.1$ and restricted to
         an angular region of $5^{\circ}$ around the actual burst location.}
\label{fig:tott2}
\end{figure}

Reduction of the background without significant signal loss can be achieved
by only investigating a certain angular region centered around the actual GRB position.
As detector angular resolution we have $\sigma_{a}=1^{\circ}$, so restricting
ourselves to an angular region of $5^{\circ}$ around the GRB location will
reduce significantly the background while preserving basically all signal muons.\\
The stacked time profile of our previous generation, but now restricted to
an angular region of $5^{\circ}$ around the burst location, is shown in Fig.~\ref{fig:tott2}.  

\begin{figure}[htb]
\begin{center}
\includegraphics[keepaspectratio,width=9cm]{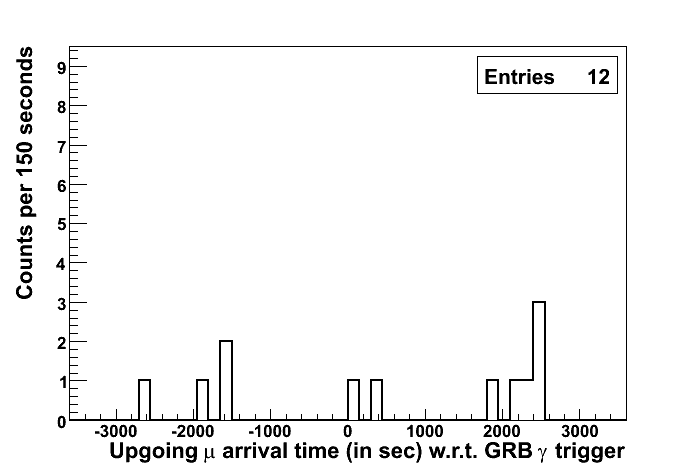}
\end{center}
\caption{Stacked time profile corresponding to the background data of Fig.~\ref{fig:tott2}.}
\label{fig:bkgt2}
\end{figure}

Visual inspection of Fig.~\ref{fig:tott2} raises some doubts to a conclusion
that the observed time profile results from a uniform background distribution.
This is confirmed if we investigate the corresponding background distribution as shown
in Fig.~\ref{fig:bkgt2}.

Comparison of Fig.~\ref{fig:tott2} and Fig.~\ref{fig:bkgt2} allows the identification
of the GRB signals in the central bin.
In the analysis of experimental data, however, we don't have access to the actual corresponding
background distribution. As such, we need to quantify our degree of (dis)belief in a background
observation solely based on the actually recorded signals like in Fig.~\ref{fig:tott2}.

\section{Bayesian assessment of the significance}
Consider two propositions $A$ and $B$ and some prior information $I$.
We introduce the notation $p(A|BI)$ to represent the probability that $A$ is true
under the condition that both $B$ and $I$ are true.
Following the arguments of extended logic \cite{jaynes} we automatically arrive
at the so-called theorem of Bayes
\begin{equation}
p(B|AI)=p(B|I)\,\frac{p(A|BI)}{p(A|I)} \quad .
\label{eq:bayes}
\end{equation}
The above theorem is extremely powerful in the process of hypothesis testing,
as will be shown here.\\
Consider a hypothesis $H$ in the light of some observed data $D$ and prior information $I$.
By $H_{\ast}$ we denote an unspecified alternative to $H$. This implies that $H_{\ast}$
is just the proposition that $H$ is false.
From eq.~\eqref{eq:bayes} we immediately obtain
\begin{equation}
\frac{p(H|DI)}{p(H_{\ast}|DI)}=\frac{p(H|I)}{p(H_{\ast}|I)}\,\frac{p(D|HI)}{p(D|H_{\ast}I)} \quad .
\label{eq:bayes2}
\end{equation}
Introducing an intuitive decibel scale, we can express the evidence $e(H|DI)$ for $H$ relative to
any alternative based on the data $D$ and prior information $I$ as~: 
\begin{equation}
e(H|DI) \equiv 10\log_{10} \left[\frac{p(H|DI)}{p(H_{\ast}|DI)} \right] \quad .
\label{eq:evidence}
\end{equation}
Combined with eq.~\eqref{eq:bayes2} this yields
\begin{equation}
e(H|DI)=e(H|I)+10\log_{10} \left[\frac{p(D|HI)}{p(D|H_{\ast}I)} \right] \quad .
\label{eq:evidence2}
\end{equation}

To quantify the degree to which the data support a certain hypothesis $H$, we introduce
the Bayesian observables $\psi \equiv -10\log_{10} p(D|HI)$ and
$\psi_{\ast} \equiv -10\log_{10} p(D|H_{\ast}I)$.
Since the value of a probability always lies between 0 and 1, we have $\psi \geqq 0$
and $\psi_{\ast} \geqq 0$. Together with eq.~\eqref{eq:evidence2} we obtain
\begin{equation}
e(H_{\ast}|DI)=e(H_{\ast}|I)+\psi-\psi_{\ast} \leqq e(H_{\ast}|I)+\psi \quad .
\label{eq:evidence3}
\end{equation}
In other words : there is no alternative to a certain hypothesis $H$ which can be supported by the
data $D$ by more than $\psi$ decibel, relative to $H$.\\
So, the value $\psi=-10\log_{10} p(D|HI)$ provides the reference to quantify our
degree of belief in $H$.

In our evaluation of the stacked time profile the main question is to which degree we
believe our observed distribution to be inconsistent with respect to a uniform background.
This question can be answered unambiguously if we are able to determine the $\psi$
value corresponding to the uniform background hypothesis based on our observed stacked time profile.

The process of recording background signals is identical to performing an experiment
with $m$ different possible outcomes $\{A_{1},...,A_{m}\}$ at each trial.
Obviously, $m$ is in our case just the number of bins in the time profile and the
number of trials $n$ is the number of entries.\\
In case all the probabilities $p_{k}$ corresponding to the various outcomes $A_{k}$
on successive trials are independent and stationery, the experiment is said to belong
to the Bernoulli class $B_{m}$ \cite{jaynes}.
It is clear that our data recordings according to a uniform background hypothesis
satisfy the requirements of $B_{m}$.
The probability $p(n_{1} \dots n_{m}|B_{m}I)$ of observing $n_{k}$ occurrences
of each outcome $A_{k}$ after $n$ trials is therefore given by the multinomial
distribution \cite{jaynes}.
Consequently, the probability for observing a specific set of background data $D$
consisting of $n$ entries is given by 
\begin{equation}
p(D|B_{m}I)=\frac{n!}{n_{1}! \cdots n_{m}!} \, p_{1}^{n_{1}} \cdots p_{m}^{n_{m}} \quad .
\label{eq:pbm}
\end{equation}
This immediately yields the following expression for the $\psi$ value according to 
a uniform background hypothesis
\begin{equation}
\psi=-10 \left[ \log_{10}n! + \sum_{k=1}^{m}(n_{k}\log_{10}p_{k}-\log_{10}n_{k}!) \right]~.
\label{eq:psi}
\end{equation}

When a signal from a uniform background is being recorded in our time window, there is
no preference for any specific time bin. This implies that in our case all $p_{k}$
values are identical and equal to $m^{-1}$.
As such we can evaluate the $\psi$ value of eq.~\eqref{eq:psi} for any set of observed data $D$.

\subsection{Relation to a frequentist approach}
In the case of large statistics we can use Stirling's approximation $\ln x!=x \ln x - x$
for $x \gg 1$ in eq.~\eqref{eq:psi}.
Together with the fact that $\sum n_{k}=n$ this yields the frequentist approximation
\begin{equation}
\psi=10 \sum_{k=1}^{m} n_{k}\log_{10}\left(\frac{n_{k}}{np_{k}} \right) \quad .
\label{eq:psi2}
\end{equation}
Furthermore, for a "near match" scenario we have $n_{k} \approx np_{k}$.
In such a case we can use the series expansion $\ln x=(x-1)-\frac{1}{2}(x-1)^{2}+\dots$,
which yields
\begin{equation}
\left| \sum_{k=1}^{m} n_{k}\ln\left(\frac{n_{k}}{np_{k}} \right) \right|
 \approx \frac{1}{2} \sum_{k=1}^{m} \frac{(n_{k}-np_{k})^{2}}{np_{k}} \quad .
\label{eq:match}
\end{equation}
This yields the correspondence with the $\chi^{2}$ statistic
\begin{equation}
\chi^{2}=\sum_{k=1}^{m} \frac{(n_{k}-np_{k})^{2}}{np_{k}} \quad .
\label{eq:chi2}
\end{equation}

Equation~\eqref{eq:chi2} allows a frequentist $\chi^{2}$ evaluation of the statistical
significance of our observations. However, this will only provide meaningful results
in case the conditions mentioned above are satisfied. In case a rather unlikely event
happens to be observed within a small number of trials, a $\chi^{2}$ analysis may lead
to completely wrong conclusions whereas the Bayesian approach outlined above will
provide the correct results \cite{jaynes}.
As such, the present studies will be based on the exact Bayesian expression of eq.~\eqref{eq:psi}. 

\section{Discovery potential}
Evaluation of the expression of eq.~\eqref{eq:psi} for the data displayed in Fig.~\ref{fig:tott1}
yields $\psi=713.38$~dB. Since these data don't allow the identification of a GRB signal,
this rather high $\psi$ value must be due to background fluctuations.
This is indeed confirmed by investigation of the corresponding background data shown
in Fig.~\ref{fig:bkgt1}, which yield $\psi_{bkg}=709.43$~dB.
Consequently, it is required to determine the $\psi$ value of the corresponding background
before the statistical significance of an observed time profile can be evaluated.

In our toy model studies we have directly access to the corresponding background time profile,
but this will in general not be the case in an actual experimental data analysis effort.
One way to investigate background signals is to record data as outlined above, but with
fictative GRB trigger times not coinciding with the actual $t_{grb}$.
This method we call "on source off time". In order to have similar detector conditions
for both the signal and background studies, the fictative trigger times should be chosen
not too distinct from the actual $t_{grb}$.
Recording background data in a time span covering 1 day before and 1 day after the
GRB observation will allow the investigation of at least 25 different background time profiles
per burst. These in turn will yield the corresponding different stacked background time profiles
which allow the determination of an average value $\bar{\psi}_{bkg}$ and the corrresponding
root mean square deviation $s_{bkg}$.

The processing of extra background data as described above might turn out to become unpractical,
due to e.g. data volume. In such a case one might consider using the remaining data of the   
off source locations of the actual time windows. Such a method we call "off source on time".\\
The performance of the "off source on time" method obviously depends on various detector
conditions, which may limit the feasibility of such a background determination.
To overcome these possible limitations, one could also envisage using the actual observed
time profile and randomise the entries in time. By performing several randomisations,
a representation of the corresponding background is obtained. This method we call "time shuffling".
It should be noted, however, that in the case of a large signal contribution the time shuffling
method will underestimate the significance of the signal.

In view of the above, we will use the "on source off time" method in our toy model studies by
generating 25 different background samples and performing our analysis procedure for
each of them.\\
In the case of the situation reflected by Fig.~\ref{fig:tott1} this yields $\bar{\psi}_{bkg}=692.04$~dB
and $s_{bkg}=21.19$~dB, which is seen to be in excellent agreement with the actual background
value corresponding to Fig.~\ref{fig:bkgt1}.\\
Comparison of the actually observed $\psi$ value of 713.38~dB with the reconstructed background values
immediately shows that no significant signal is observed.\\
However, evaluation of the data corresponding to Fig.~\ref{fig:tott2} yields $\psi=218.78$~dB
with background values $\bar{\psi}_{bkg}=99.62$~dB and $s_{bkg}=23.98$~dB.
Here a statistically significant signal is obtained.

\begin{figure}[htb]
\begin{center}
\includegraphics[keepaspectratio,width=9cm]{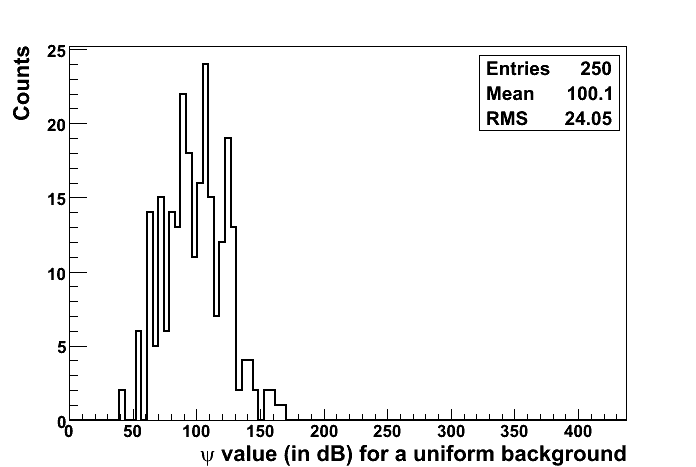}
\end{center}
\caption{Distribution of $\psi_{bkg}$ values for a large sample of background
         distributions according to the situation depicted in Fig.~\ref{fig:tott2}.}
\label{fig:psidist}
\end{figure}

For a uniform background and large statistics, the Bayesian $\psi$ observable can be approximated
by the frequentist $\chi^{2}$ statistic, as indicated in eqs.~\eqref{eq:psi2}-\eqref{eq:chi2}.
This implies that the statistical significance for deviation from a uniform background distribution
can be expressed in terms of a standard deviation $\sigma$ by comparison of the actually observed
$\psi$ value of the stacked time profile with the corresponding $\bar{\psi}_{bkg}$ and $s_{bkg}$
background values.
This is illustrated in Fig.~\ref{fig:psidist} for a sample of 250 different background samples
according to the situation reflected in Fig.~\ref{fig:tott2}.
The distribution of the obtained $\psi_{bkg}$ values as shown in Fig.~\ref{fig:psidist}
exhibits a Gaussian profile with a mean value and standard deviation
which are indeed consistent with the above $\bar{\psi}_{bkg}$ and $s_{bkg}$ values, respectively.

\begin{figure}[htb]
\begin{center}
\includegraphics[keepaspectratio,width=9cm]{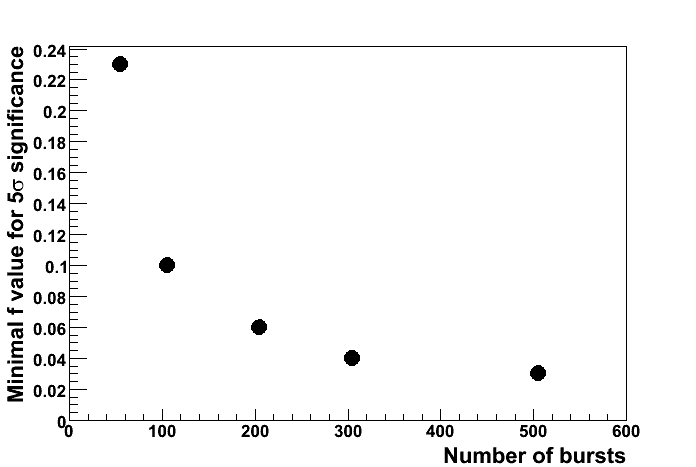}
\end{center}
\caption{Sensitivities corresponding to a $5\sigma$ signal significance.}
\label{fig:disc}
\end{figure}

Variation of the number of GRBs allows a determination of the minimal value of the fraction $f$
for which a statistically significant signal can be obtained.
Common practice is to claim a discovery in the case a significance in excess of $5\sigma$
is obtained. Following the procedure outlined above this leads to the discovery sensitivities
as shown in Fig.~\ref{fig:disc}.

It should be noted that the actually achievable sensitivities are depending on various
detector specific parameters and the quality of the available data.
As such, all parameters as well as the amount of possible background samples will have to be
optimised for each specific experimental data analysis scenario.  

In case no significant signal can be identified from an experimentally observed stacked time profile,
values like the ones presented in Fig.~\ref{fig:disc} provide the basis for a fluence limit
determination.

\section{Summary}
The method introduced in this report allows identification of high-energy neutrinos
from gamma ray bursts with large scale neutrino telescopes.
The procedure is based on a time profile stacking technique, which provides statistical
significant results even in the case of low signal rates.

The performance of the method has been investigated by means of toy model studies based on
realistic parameters for the future IceCube km$^{3}$ neutrino telescope and a variety of burst samples.
From these investigations it is seen that a $5\sigma$ significance is obtained on a
sample of 500 bursts with a signal rate as low as 1 detectable neutrino for 3\% of the bursts.\\
Finally, it should be realised that the actually achievable sensitivities 
are depending on various detector specific parameters and the quality of the available data.
These aspects, however, fall beyond the scope of the present report.

\begin{ack}
The author would like to thank Bram Achterberg, Martijn Duvoort, John Heise and Garmt de Vries
for the very fruitful discussions on the subject.
\end{ack}

\end{document}